# Reduced Variability in Threshold Switches Using Heterostructures of SiO$_x$ and Vertically Aligned MoS$_2$

Jimin Lee[1], Rana Walied Ahmad[2], Sofía Cruces[1], Dennis Braun[1], Lukas Völkel[1], Ke Ran[3,4,5], Joachim Mayer[4,5], Stephan Menzel[2], Alwin Daus[1,6], and Max C. Lemme[1,3,*]

[1] Chair of Electronic Devices, RWTH Aachen University, Otto-Blumenthal-Str. 25, 52074 Aachen, Germany.

[2] JARA-Fit and Peter Grünberg Institute (PGI-7), Forschungszentrum Jülich GmbH, Wilhelm-Johnen-Str., 52428 Jülich, Germany.

[3] AMO GmbH, Advanced Microelectronic Center Aachen, Otto-Blumenthal-Str. 25, 52074 Aachen, Germany.

[4] Central Facility for Electron Microscopy, RWTH Aachen University, Ahornstr. 55, 52074 Aachen, Germany.

[5] Ernst Ruska Centre for Microscopy and Spectroscopy with Electrons, Forschungszentrum Jülich GmbH, Wilhelm-Johnen-Str., 52428 Jülich, Germany.

[6] Institute of Semiconductor Engineering, University of Stuttgart, Pfaffenwaldring 47, 70569 Stuttgart, Germany.

\* Corresponding author: Max C. Lemme

E-mail address: max.lemme@eld.rwth-aachen.de

**Abstract**

Layered two-dimensional (2D) materials provide unique structural features, such as physical gaps between their layers that are only connected through van der Waals (vdW) forces. These vdW gaps can guide the migration of intercalated ions and thus regulate filament growth in resistive switching (RS) devices. Vertically aligned 2D materials and their heterostructures provide vdW gap-mediated ion transport in memristor crossbars, providing great potential for high-density integration and reliable RS performance. Nevertheless, the fundamental switching mechanisms and their contributions to the RS remain inadequately understood. In this work, we investigate silver (Ag) filament-based threshold switching (TS) in heterostructures comprising vertically aligned 2D molybdenum disulfide (VAMoS$_2$) grown via sulfurization and silicon oxide (SiO$_x$). Compared to SiO$_x$-only devices, the SiO$_x$/VAMoS$_2$ devices exhibit TS with higher on-threshold and hold voltages, each approximately 0.4 V, faster switching times down to 356 ns under a 4 V pulse, and a lower cycle-to-cycle on-current variability of 3.0%. A physics-based, variability-aware model reveals that confined Ag ion migration within the vdW gaps in VAMoS$_2$ forms ultrathin seed filaments, which guide filament growth in the SiO$_x$ layer. These findings establish SiO$_x$/VAMoS$_2$ heterostructures as a promising concept for reliable TS in vertical device architectures for emerging memories and neuromorphic computing.

# 1. Introduction

Two-dimensional materials (2DMs) have gained attention as resistive switching (RS) media for neuromorphic and in-memory computing applications[1–4]. Various 2DMs, including transition metal dichalcogenides (TMDs)[5–7], hexagonal boron nitride (h-BN)[8–10], 2D heterostructures[11–13], and oxide-2D heterostructures[14–17], have shown memristive RS characteristics. 2DMs can be stacked vertically with other materials without requiring lattice matching[18], which enables multifunctional three-dimensional integration[19]. This capability opens opportunities for compact, high-performance memory and computing technologies based on 2D material memristive devices[6,20].

Two-terminal memristive devices based on 2DMs, consisting of a metal-2DM-metal configuration, are particularly attractive for energy-efficient technologies because of their small footprint, high integration density, and low switching energy[1,5]. Recently, vertically aligned 2DMs have emerged as promising RS media for such vertical two-terminal devices[19,21–25]. For example, vertically aligned h-BN grown by plasma-enhanced chemical vapor deposition has shown nonvolatile RS with a high device yield of ~90%, multistate conductance, and compatibility with complementary metal oxide semiconductor technology[19]. Vertically aligned molybdenum disulfide (VAMoS$_2$) grown by chemical vapor deposition (CVD) has shown volatile RS with an on/off ratio of ~10$^6$, enabling artificial neuron emulation[23]. Nonvolatile RS has also been observed in VAMoS$_2$ on silicon (Si) with chromium (Cr)/gold (Au) electrodes, where switching is caused by hydroxyl ions originating from the catalytic splitting of adsorbed water molecules[21]. The heterostructures of VAMoS$_2$ and silicon oxide (SiO$_x$) have further demonstrated silver (Ag)-based filamentary threshold RS with endurance exceeding 10$^4$ cycles[26]. These findings highlight the potential of vertically aligned 2D materials for use in memristive devices; however, their switching mechanisms or their roles in the RS behavior of heterostructures remain insufficiently understood, e.g., on the influence of the vertical alignment of 2DMs or their presence in heterostructures on key performance parameters such as the switching speed, variability, and reliability.

Memristive devices change their electrical resistance in response to an applied voltage[27]. One of the predominant RS mechanisms is electrochemical-metallization (ECM)[28,29], where metallic filaments form and dissolve within the switching medium due to the migration of active metal ions such as Ag, copper (Cu), or nickel (Ni)[11,30–32]. The formation and dissolution of conductive filaments (CFs) are inherently stochastic, controlled by the electric field, the switching medium, and the electrode interface, all of which influence filament stability and morphology[9,11,33]. Thick and bulky CFs typically result in nonvolatile RS, whereas thin and fragile CFs rupture spontaneously after bias removal, leading to volatile RS, also known as threshold switching (TS), observed in diffusive memristors[28,33,34]. In 2DMs, ECM-based RS is driven by ion migration along intrinsic structural features such as van der Waals (vdW) gaps, grain boundaries, and defects[21,22,35–41]. Since 2DMs play a crucial role in determining the switching dynamics, material selection and device configuration must be considered when optimizing the device performance.

In this work, we compare Ag filament-based TS in vertical memristive devices consisting of $SiO_x$ with and without $VAMoS_2$ layers. We elucidate the influence of the $VAMoS_2$ layers on the TS behavior, with a particular focus on the switching dynamics based on filament formation, by combining experimental data with a physics-based JART ECM simulation model[42–44]. The presence of the $VAMoS_2$ layer regulates Ag ion migration through its van der Waals (vdW) gaps, leading to confined filament formation with faster switching speeds and slightly higher average switching voltages with significantly lower cycle-to-cycle variability compared to devices based on $SiO_x$ alone.

## 2. Results and Discussion

### 2.1. Device fabrication and structure

Our devices consist of a vertical stack with an active Ag top electrode (TE) and an inert Au bottom electrode (BE). The switching medium is either a $SiO_x$/VAMoS$_2$ heterostructure or a $SiO_x$ layer ($SiO_x$-only), as shown in Figures 1a and 1b. These configurations allow us to investigate the influence of the VAMoS$_2$ layer on the RS behavior during CF formation.

Device fabrication began with the patterning of Au BEs on a Si substrate covered with a 300 nm thick thermal $SiO_2$ film using photolithography, e-beam evaporation, and lift-off. For heterostructure devices, a 6 nm molybdenum (Mo) film was then sputtered and patterned on the predefined Au Bes. The VAMoS$_2$ layer was synthesized by sulfurizing the Mo film in a sulfur atmosphere at 800 °C[26,45]. During this process, an amorphous $SiO_x$ layer formed on the VAMoS$_2$ surface by diffusion of Si and O from the substrate[26,46,47]. For $SiO_x$-only devices, a comparable $SiO_x$ layer was formed on the Au BEs without a Mo film but otherwise under identical fabrication conditions. Finally, the Ag TE, capped with an Au layer, was patterned via photolithography, e-beam evaporation, and lift-off. Detailed descriptions of the device fabrication and material growth processes are provided in Section 4.1 of the Methods section.

Cross-sectional transmission electron microscopy (TEM) confirmed the device stacks (see Figure S1 in the Supplementary Information (SI)). Energy-dispersive X-ray spectroscopy (EDXS) elemental mapping verified the presence of Si, O, Mo, and S in the heterostructure between the Ag TE and the Au BE (Figure 1c) and comparable Si and O signals in the $SiO_x$-only device (Figure 1d). High-resolution TEM (HRTEM) imaging revealed the vertical alignment of the MoS$_2$ layers in the $SiO_x$/VAMoS$_2$ device (Figure 1e). Raman spectroscopy confirmed the crystalline 2H-phase of MoS$_2$, with characteristic peaks at 382 cm$^{-1}$ and 407 cm$^{-1}$, corresponding to the E$^1_{2g}$ and A$_{1g}$ modes, respectively (Figure 1f)[48]. The E$^1_{2g}$/A$_{1g}$ peak intensity ratio of ~0.4 further supported the vertical orientation of MoS$_2$[49].

## 2.2. Direct-current (dc) characterization

Direct-current (dc) current–voltage (*I*–*V*) measurements were performed on SiO$_x$/VAMoS$_2$ and SiO$_x$-only devices by applying a bias voltage to the Ag TE while grounding the BE. A current compliance ($I_{cc}$) of 1 μA was set for the measurements. Both devices exhibited reproducible TS over 128 consecutive cycles without requiring an initial electroforming step (Figures 2a and 2b). The SiO$_x$/VAMoS$_2$ device showed gradual current transitions, whereas the SiO$_x$-only device showed abrupt transitions. During the forward sweep, both devices switched from the initial high-resistance state (HRS) to the low-resistance state (LRS) at the on-threshold voltage ($V_{t,on}$). In the reverse sweep, the devices remained in the LRS until the voltage dropped below the hold voltage ($V_{hold}$), after which they returned to the HRS. The off-threshold voltage ($V_{t,off}$) was defined as the voltage at which the current returned to the initial HRS current level.

We extracted the switching voltage parameters ($V_{t,on}$, $V_{hold}$, and $V_{t,off}$) from 128 consecutive switching cycles of the SiO$_x$/VAMoS$_2$ and SiO$_x$-only devices. The distributions of the parameters are shown as histograms and cycle-dependent plots in Figures S2 and S3 (Supplementary Information). Compared with the SiO$_x$-only devices, the SiO$_x$/VAMoS$_2$ devices showed smaller fluctuations in all three voltage parameters. For example, the $V_{t,on}$ values of the SiO$_x$/VAMoS$_2$ device ranged from 0.3 V to 0.5 V, whereas those of the SiO$_x$-only device varied widely between 0.1 V and 0.6 V. The normalized cumulative distribution functions (CDFs) of $V_{t,on}$, $V_{hold}$, and $V_{t,off}$ over 128 cycles show that the presence of the VAMoS$_2$ layer increases the average switching voltage while reducing the cycle-to-cycle variability (Figure 2c). The vertically aligned vdW gaps in the VAMoS$_2$ layers provide predefined ion migration pathways, which may promote confined and reproducible formation and rupture of the CFs. The average $V_{hold}$ also increased from $0.1 \pm 0.04$ V in SiO$_x$-only to $0.4 \pm 0.02$ V in the SiO$_x$/VAMoS$_2$ heterostructure, indicating easier rupture of the CFs in the heterostructure, which is potentially due to a structurally weak point at the SiO$_x$/VAMoS$_2$ interface. The devices returned to their initial HRS at $V_{t,off}$ of $\sim 0.1 \pm 0.02$ V for the heterostructure devices and $\sim 0.04 \pm 0.04$ V for the SiO$_x$-only devices.

Figures 2d and 2e show the TS in both device types for different $I_{cc}$ values, i.e., over four orders of magnitude from 10 nA to 100 μA. The SiO$_x$/VAMoS$_2$ device exhibited stepwise switching, with an initial abrupt current increase followed by a gradual rise (Figure 2d), whereas the SiO$_x$-only device showed sharp and abrupt transitions once $V_{t,on}$ was reached (Figure 2e). While the origin of the stepwise switching in the SiO$_x$/VAMoS$_2$ device is not yet understood and requires further investigation, the high on-state current and reproducible TS behavior suggest its potential for selector applications[50,51] and neuromorphic computing, particularly for implementing artificial neurons[23].

## 2.3. Pulsed voltage characterization

Pulsed voltage measurements were conducted to explore the transient switching dynamics. The time-dependent current response was recorded under voltage pulses of varying amplitudes ($V_{pulse}$). As shown in Figure 3a, the switching time ($t_{on}$) corresponds to the time required for the current to reach 90% of the on-current ($I_{on}$)[26,52]. For each measurement, a voltage pulse was applied to switch the device from the HRS to the LRS within $t_{on}$, after which the device spontaneously returned to the HRS. The HRS level was quantified as the off-current ($I_{off}$) by using a subsequent read pulse of 0.3 V ($V_{read}$) with a duration of 3 μs. The on- and off-switching characteristics are described in detail in the literature[26,52]. Figure 3a shows a representative transient response of the SiO$_x$/VAMoS$_2$ device to a 4 V pulse with a duration of 2 μs. The inset presents the corresponding applied voltage pulse waveform.

We plotted the $I_{on}$ and $I_{off}$ values from 30 consecutive cycles of a $V_{pulse}$ of 4 V for 2 μs for a SiO$_x$/VAMoS$_2$ and a SiO$_x$-only device (Figures 3b and 3c). The heterostructure device had significantly lower variations in $I_{on}$ (3.0%) compared to the SiO$_x$-only device (46.5%). The $t_{on}$ values were fitted with Gaussian distributions with mean values (μ) and standard deviations (σ) of 620 ns and 160 ns for the SiO$_x$/VAMoS$_2$ and 940 ns and 450 ns for the SiO$_x$-only device (Figure 3d). The broader $t_{on}$ distribution of the SiO$_x$-only devices reflects higher variability. Given that the switching dynamics can be modulated by the voltage amplitude[11,29,33] and that $t_{on}$ is a critical parameter in memristive device operation[53,54], we investigated the dependence of $t_{on}$ on $V_{pulse}$ by applying 30 consecutive pulses. Compared with the SiO$_x$-only device, the SiO$_x$/VAMoS$_2$ device demonstrated relatively faster $t_{on}$ values with reduced variations across amplitudes of 4 V, 4.5 V, and 5 V (Supplementary Figure S4). Next, we plotted the on-state resistance ($R_{on}$) as a function of $V_{pulse}$ for both device types (Figure 3e). Compared with the SiO$_x$-only devices, the SiO$_x$/VAMoS$_2$ devices exhibited lower average $R_{on}$ values with reduced cycle-to-cycle variability. These results suggest that the vertically aligned vdW gaps in the VAMoS$_2$ confine ion migration[21,22,39,40], leading to controlled filament growth and rupture, faster switching speeds, and reduced variability. Furthermore, the $R_{on}$ values decreased with increasing amplitude of $V_{pulse}$ for both devices, which is consistent with stronger Ag ion drift, enhanced CF formation, and increased current flow at higher pulse amplitudes[55]. In Section 2.4, we further discuss the switching variability and filament formation mechanisms by correlating the experimentally observed data with the physics-based JART ECM simulation model[42–44] and by comparing devices with and without the VAMoS$_2$ layer.

## 2.4. Conductive filament formation and variability modeling

Figure 4a shows schematic cross-sections of our proposed CF formation process in SiO$_x$/VAMoS$_2$ and SiO$_x$-only devices in the LRS. In both devices, applying a positive bias to the Ag TE induces oxidation of Ag, generating Ag ions that migrate toward the Au BE. Upon reaching the BE, the Ag ions are reduced and form a metallic nucleation seed at the interface between the switching medium and the BE[43,56]. The seed initiates the growth of the CFs upward from the BE toward the TE, switching the device from the HRS to the LRS[43]. In the SiO$_x$/VAMoS$_2$ devices, the vdW gaps in the VAMoS$_2$ layers function as ion-conducting channels that confine Ag ion migration[22,39,40], leading to the nucleation of ultrathin seed filaments at the BE and their subsequent extension along the vdW gaps. The VAMoS$_2$ layer thus directs and restricts filament growth. At the SiO$_x$/VAMoS$_2$ interface, the pre-confined seed filaments provide favorable points that promote extension into the overlying SiO$_x$ layer, where the seed filaments can drive the development of thicker filaments in the SiO$_x$ layer. In contrast, the SiO$_x$-only devices lack such guided pathways, which leads to randomly initiated thin filaments that form directly from the Au BE. Notably, in both device types, the CFs remain sufficiently thin to rupture spontaneously, enabling TS.

Figures 4b and 4c show the transient current responses of a SiO$_x$/VAMoS$_2$ and a SiO$_x$-only device, respectively, under a $V_{pulse}$ = 4 V for a duration of 2 μs, which was repeated for 30 cycles. These experimental data confirm faster and more consistent switching in the SiO$_x$/VAMoS$_2$ heterostructure than in the SiO$_x$-only device. A physics-based, variability-aware ECM model[43] reproduced the experimental data for the SiO$_x$/VAMoS$_2$ and SiO$_x$-only devices via a cycle-to-cycle variability fit, as shown in Figures 4d and 4e. Details of the model are provided in Section 4.4 of the Methods section, and the fitting parameters are summarized in Table S1 in Supplementary Section S5. The model achieved good agreement with the experimental data, successfully capturing the faster $t_{on}$ and narrower $t_{on}$ distributions observed in the $SiO_x$/VAMoS$_2$ devices.

The simulations suggest that the variability originates from the filament radius ($r_{fil}$) and the stochastic distance ($x$) between the filament tip and the TE. In heterostructure devices, the interlayer gap of the VAMoS$_2$ layers forms ultrathin filaments, estimated to be only 1–2 vdW gaps wide. The gap spacing of ~6.5 Å[22,57] exceeds the Ag atomic diameter of 3.44 Å[58], which allows confined ion migration. The filament in VAMoS$_2$ is restricted and mainly experiences vertical growth along a vdW gap[22,39]. Hence, it is assumed that the filament does not substantially grow in the horizontal direction within VAMoS$_2$ after nucleation. Thus, the diameter of the ultrathin filament within VAMoS$_2$ is as small as the diameter of the nucleus at the VAMoS$_2$/Au interface. These ultrathin filaments in VAMoS$_2$ ($r_{fil(1,VAMoS2)}$) subsequently extend into the $SiO_x$ layer, where they seed the growth of thicker filaments in the $SiO_x$ layer ($r_{fil(1,SiOx)} > r_{fil(1,VAMoS2)}$). In the $SiO_x$-only devices, filaments nucleate randomly at the Au/$SiO_x$ interface with radii larger than those in the VAMoS$_2$ region but smaller than those in the $SiO_x$ region of the heterostructure devices ($r_{fil(1,SiOx)} > r_{fil(2,SiOx)} > r_{fil(1,VAMoS2)}$, see Figure 4a). In the $SiO_x$-only devices, the subsequent filament diameter in $SiO_x$ is not coupled to the nucleus diameter at the Au/$SiO_x$ interface and grows with larger diameters.

Thus, the pre-confined seeds at the $SiO_x$/VAMoS$_2$ interface play a critical role in initiating thicker and more reproducible filament growth in the $SiO_x$ layers of the heterostructures. The cycle-to-cycle variability model further showed that the filament radius is closely linked to the stochastic parameter $\delta$, describing the stochastic change in the distance $x$ between the filament tip and TE. A larger filament corresponds to smaller variability and to a smaller $\delta$ because the addition or removal of single atoms does not strongly affect the filament geometry and the current through it. Conversely, a narrow filament displays higher variability and a larger $\delta$, as single-atom changes significantly alter the filament geometry and the electronic current, according to the exponential relation of $x$ in the electronic tunneling current. Owing to stochasticity, discrete current jumps occur as a few atoms are added or removed from the filament tip at random intervals, modifying the minimum distance between the filament tip and TE and the current response.

We attribute the improved TS performance to the VAMoS$_2$ layer directing thin CF growth and providing a structurally weak point at the SiO$_x$/VAMoS$_2$ interface, where it easily breaks. Our results align with previous findings where restricting Ag ion transport through atomic defects in graphene reduced the number of ions entering the switching medium[11].

## 3. Conclusion

We compared Ag filament-based TS in vertical, two-terminal devices comprising SiO$_x$/VAMoS$_2$ and SiO$_x$ as switching layers. Compared with the SiO$_x$-only devices, the SiO$_x$/VAMoS$_2$ devices showed enhanced switching characteristics, including faster switching times, higher $V_{hold}$ with significantly reduced variability, and more stable and repeatable transient current responses. Our study suggests that these improvements originate from the vdW gaps in the VAMoS$_2$ layers, which guide and confine Ag ion migration. This confinement enables the formation of ultrathin seed filaments that extend into the overlying SiO$_x$, where they initiate stable conductive pathways. The combination of electrical and analytical experiments with physics-based ECM simulations confirms that the filament morphology determines the switching dynamics and variability. In particular, the pre-confined seed filaments at the SiO$_x$/VAMoS$_2$ interface establish a well-controlled environment for filament growth, leading to more reliable TS behavior. These findings show that vertically aligned 2D materials offer an effective strategy for engineering CF formation in memristive devices and suggest their strong potential in emerging memory and neuromorphic computing applications.

## 4. Methods

### 4.1 Device fabrication and material growth

Devices with lateral dimensions of 4 μm x 4 μm were fabricated on Si chips (2 cm x 2 cm) covered with 300 nm thermal $SiO_2$. The bottom electrodes (BEs) were patterned via a double resist stack (LOR 3A/AZ5214E from MicroChemicals GmbH/Merck Performance Materials GmbH) and optical contact lithography with an EVG 420 mask aligner. A 5 nm Ti adhesion layer and a 50 nm Au layer were deposited via electron-beam evaporation in a Pfeiffer tool and lifted off in dimethyl sulfoxide (DMSO) heated to 80 °C. The Mo films were patterned via the same double resist stack and lithography process. A 6 nm Mo layer was deposited via dc sputtering with the CREAVAC tool (CREAMET S2) and lifted off in DMSO. For $SiO_x$-only devices, Au BEs without Mo patterns were retained on the same chip. $SiO_x$/VAMoS$_2$ and $SiO_x$ stacks were synthesized on Mo/Au BEs and Au BEs, respectively, by sulfurization in a CVD tube furnace (RS 80/300, Nabertherm GmbH). Vertically aligned MoS$_2$ (VAMoS$_2$) layers were formed by thermally assisted conversion of Mo in a sulfur atmosphere at 800 °C for 30 min with an Ar carrier gas (20 sccm) at a pressure of $9.6 \times 10^{-2}$ mbar[26]. Sulfur diffusion converted the Mo film into ~20 nm thick VAMoS$_2$ layers, while ~10 nm amorphous $SiO_x$ simultaneously formed on the VAMoS$_2$ and Au BE surfaces[26,46]. Finally, the top electrodes (TEs) were defined via the same lithography process as the BE. A 30 nm

Ag layer capped with a 30 nm Au layer was deposited via electron-beam evaporation, followed by lift-off.

### 4.2 Material characterization

Transmission electron microscopy (TEM) analysis was carried out to investigate the cross-sectional device structure, the orientation of the $MoS_2$ layers, and the chemical composition. TEM lamellae were prepared via a focused ion beam (FIB) in an FEI Strata400 system. TEM imaging and energy-dispersive X-ray spectroscopy (EDXS) were conducted with a JEOL JEM F200 instrument operated at 200 kV. Raman measurements were conducted via a WITec alpha 300R Raman system with a 532 nm excitation laser operating at 1 mW power and an 1800 g/mm grating.

### 4.3 Electrical characterization

Electrical measurements were conducted under ambient conditions at room temperature using a Lakeshore probe station connected to a semiconductor parameter analyzer (Keithley 4200S-SCS, Tektronix) equipped with two source measure unit (SMU) cards (Keithley 4200-SMU) and preamplifiers (Keithley 4200-PA). A voltage bias was applied to the Ag TE, and the Au BE was grounded. The applied voltage was monitored from the Ag TE, and the output current was recorded from the Au BE. For dc voltage sweeps, the voltage was applied from 0 V to 1.0 V (forward sweep, labeled '1') and returned to 0 V (reverse sweep, labeled '2') with a sweep rate of 0.3 $Vs^{-1}$ and a step size of 0.01 V (see Figures 2a and 2b). The current compliance was controlled by an external current limiter of the Keithley 4200-SCS. Pulsed voltage measurements were performed using Keithley 4225-PMU pulse measure units.

### 4.4 Simulation model

The JART ECM v1 model[59] is a deterministic compact model. It was used to capture the ECM device switching by tracking the effective tunneling distance $x$ between the active electrode (TE) and the filament tip, which is modulated by filament growth[42,43]. The device was represented by two parallel branches between the active TE and inert BE: (i) an electronic tunneling path $I_{Tu}$ and (ii) an ionic path $I_{ion}$ that comprised nucleation, redox (electron-transfer) reactions at both interfaces, and ion migration through the switching medium.

The state evolution, i.e., the evolution of the tunnelling distance $x$, is described by Faraday's law[60]:

$$\frac{dx}{dt} = -\frac{M_{Me}}{ze\rho_{m,Me}} j_{ion}, \qquad (1)$$

where $j_{ion}$ is the ionic current density, $M_{Me}$ is the atomic mass of the deposited active electrode's metal, $z$ is the charge transfer number, and $\rho_{m,Me}$ is its mass density. The nucleation time, which is the time needed to build a pre-filament seed at the inert electrode/switching medium interface, is given as[61,62]:

$$t_{nuc} = t_{0,nuc} \exp\left(\frac{\Delta G_{nuc}}{k_B T}\right) \exp\left(-\frac{(N_c + \alpha_{nuc})ze}{k_B T} \eta_{nuc}\right). \qquad (2)$$

Here, $t_{0,nuc}$ denotes the nucleation time prefactor, $\Delta G_{nuc}$ is the nucleation activation energy, $k_B$ is the Boltzmann constant, $T$ is the temperature, $N_c$ is the number of metal atoms needed to achieve a stable nucleation seed, $\alpha_{nuc}$ is the charge transfer coefficient for nucleation, and $\eta_{nuc}$ is the nucleation overpotential. The electron-transfer (redox) processes are given by the Tafel equation[63], a simplified version of the Butler-Volmer equation. At the filament/insulator interface during switching, this equation is as follows:

$$I_{\text{fil}} = j_{0,\text{et}} A_{\text{fil}} \left( \exp\left(-\frac{\alpha_{\text{et}} ez}{k_B T} \eta_{\text{fil}}\right) - 1 \right), \qquad (3)$$

whereas at the active electrode/insulator interface during the switching, it reads:

$$I_{\text{ac}} = j_{0,\text{et}} A_{\text{ac}} \left( \exp\left(\frac{(1-\alpha_{\text{et}}) ez}{k_B T} \eta_{\text{ac}}\right) - 1 \right). \quad (4)$$

The respective electron transfer overpotentials are given as $\eta_{\text{fil}}$ and $\eta_{\text{ac}}$, where $\alpha_{\text{et}}$ is the charge transfer coefficient. The introduced areas were defined as $A_{\text{fil}} = \pi r_{\text{fil}}^2$ and $A_{\text{ac}} = \pi r_{\text{ac}}^2$ (with $r_{\text{fil}}$ the filament radius and $r_{\text{ac}}$ the active electrode radius). The exchange current density in equations (4) and (5) is portrayed as:

$$j_{0,\text{et}} = zeck_{0,\text{et}} \exp\left(-\frac{\Delta G_{\text{et}}}{k_B T}\right), \qquad (5)$$

where the electron transfer activation barrier is denoted as $\Delta G_{\text{et}}$, the ion concentration close to the interface is $c$, and the rate constant is $k_{0,\text{et}}$. The ion migration through the switching medium is described by the Mott–Gurney hopping:

$$I_{\text{hop}} = j_{0,\text{hop}} A_{\text{is}} \sinh\left(\frac{aze}{2k_B T} \frac{\eta_{\text{hop}}}{x}\right), \qquad (6)$$

with:

$$j_{0,\text{hop}} = 2zecaf \exp\left(-\frac{\Delta G_{\text{hop}}}{k_B T}\right) \qquad (7)$$

$a$ is the mean hopping distance, $f$ the hopping attempt frequency, $\Delta G_{\text{hop}}$ the migration barrier, $A_{\text{is}}$ the ionic hopping cross-section area (with $r_{\text{is}}$ the corresponding radius), and $E = \eta_{\text{hop}}/x$ the electric field. The process in the electronic branch is the electronic tunneling, given by a modified Simmons equation[64]:

$$I_{\text{Tu}} = C \frac{3\sqrt{2m_{\text{eff}}\Delta W_0}}{2x} \left(\frac{e}{h}\right)^2 \exp\left(-\frac{4\pi x}{h}\sqrt{2m_{\text{eff}}\Delta W_0}\right) A_{\text{fil}} V_{\text{Tu}}, \quad (8)$$

where $C = 2.7$[42], $m_{\text{eff}} = m_r m_0$, $\Delta W_0$ is the effective tunneling barrier, $h$ is Planck's constant, $V_{\text{Tu}}$ is the voltage across the tunneling distance, $m_r$ is the relative electron mass, and $m_0$ is the electron's remaining mass.

To model the device's cycle-to-cycle variability, two features were applied within the simulation framework[43,44]: (i) a simplified version of a previously presented cycle-to-cycle variability model and (ii) stochasticity simulation, both of which are presented in the following.

The cycle-to-cycle variability has been imposed on the parameter filament radius $r_{\text{fil}}$; thus, this value differs from cycle to cycle. The corresponding simplified variability model was applied as follows[43]: Before initiating the variability simulation, this model parameter was assigned a truncated Gaussian distribution with a certain standard deviation $r_{\text{fil},\sigma}$ centered at a defined mean $r_{\text{fil}}$. Unlike a conventional Gaussian, which allows improbable extreme values, the distribution was truncated by global physical boundaries, $r_{\text{fil,trunc,up}}$ and $r_{\text{fil,trunc,low}}$, to ensure realistic parameter ranges. A filament radius value $r_{\text{fil},1}$ was drawn randomly from the truncated distributions for one cycle. During SPICE-level simulations, 30 such values ($r_{\text{fil},1}$ - $r_{\text{fil},30}$), one for each cycle, were generated in MATLAB and stored in a text file. This file was imported by the circuit simulator (Spectre), which assigned the tabulated values directly to the compact model written in Verilog-A.

Noise was directly incorporated at each simulation step to account for stochastic effects in the state variable $x$[43]. While the deterministic update follows Faraday's law [eq. (1)], an additional stochastic term was added immediately after each time step solution. The random update interval was chosen simultaneously. This led to (the tunnelling distance's length at step $k+1$)

$$x_{k+1} = x_k - \frac{M_{Me}}{ze\rho_{m,Me}} j_{ion} \Delta t_{sim,step} \pm \delta \cdot p_{jump}, \quad (9)$$

where $\Delta t_{sim,step}$ is the simulation step size, $x_k$ is the tunnelling distance length at step $k$, and $\delta$ is the stochastic amplitude. Typically, $\delta \ll 1$ nm. This choice can be motivated by atomic dimensions: if a single Ag atom ($d_{Ag} = 0.344$ nm, $r_{Ag} = 0.172$ nm) is added or removed, the corresponding average stochastic step is

$$\delta = d_{Ag} \cdot \frac{r_{Ag}^2}{r_{fil}^2}, \quad (10)$$

with, e.g., $r_{fil}$ values of 1.22 nm or 0.8 nm, giving $\delta$ values of 0.0068 nm or 0.0159 nm, respectively. The factor $p_{jump}$ was randomly drawn between 0 and 1 and updated after $w$ steps, where $w$ itself was randomly selected in each iteration.

**Acknowledgments**

The authors acknowledge support from Dr. Marcus Hans and Prof. Jochen Schneider (StrucMatLab, RWTH Aachen University)

**Funding**

Financial support from the German Federal Ministry of Education and Research (BMBF) within the project NEUROTEC 2 (No. 16ME0398K, 16ME0399, and 16ME0400) and the Clusters4Future NeuroSys (No. 03ZU1106AA and 03ZU1106AB), by the European Union's Horizon Europe research and innovation program (via CHIPS-JU) under the project ENERGIZE (101194458), and by the Deutsche Forschungsgemeinschaft (DFG) within the project MEMMEA in SPP 2262 MemrisTec (441918103) and the Emmy Noether Programme (506140715).

**Author information**

**Authors and affiliations**

Chair of Electronic Devices, RWTH Aachen University, Otto-Blumenthal-Str. 25, 52074 Aachen, Germany: Jimin Lee, Sofía Cruces, Dennis Braun, Lukas Völkel, Alwin Daus, and Max C. Lemme

JARA-Fit and Peter Grünberg Institute (PGI-7), Forschungszentrum Jülich GmbH, Wilhelm-Johnen-Str., 52428 Jülich, Germany: Rana Walied Ahmad and Stephan Menzel

AMO GmbH, Advanced Microelectronic Center Aachen, Otto-Blumenthal-Str. 25, 52074 Aachen, Germany: Ke Ran and Max C. Lemme

Central Facility for Electron Microscopy, RWTH Aachen University, Ahornstr. 55, 52074 Aachen, Germany: Ke Ran and Joachim Mayer

Ernst Ruska Centre for Microscopy and Spectroscopy with Electrons, Forschungszentrum Jülich GmbH, Wilhelm-Johnen-Str., 52428 Jülich, Germany: Ke Ran and Joachim Mayer

Institute of Semiconductor Engineering, University of Stuttgart, Pfaffenwaldring 47, 70569 Stuttgart, Germany: Alwin Daus

**Authors' contributions**

M.C.L. conceived the research concept and designed the project framework. J.L. carried out the material synthesis, Raman analysis, device fabrication, and electrical measurements. K.R. performed the TEM and EDSX analyses. R.W.A. and S.M. provided the models and conducted the simulations. Data analysis was conducted by J.L., R.W.A., S.C., D.B., L.V., J.M., S.M., A.D., and M.C.L. All the authors contributed to the interpretation of the results. J.L. wrote the initial draft, and all the authors revised the manuscript. The work was supervised by A.D., S.M., and M.C.L.

**Corresponding author**

Correspondence to Max C. Lemme

**Availability of data and material**

The data generated and/or analyzed during this study are included in this manuscript and its supplementary material. The raw data will be made available upon reasonable request.

**Declaration**

**Competing interests**

The authors declare that they have no competing interests.

**Fig. 1.** Device Structures and Materials Characterization.

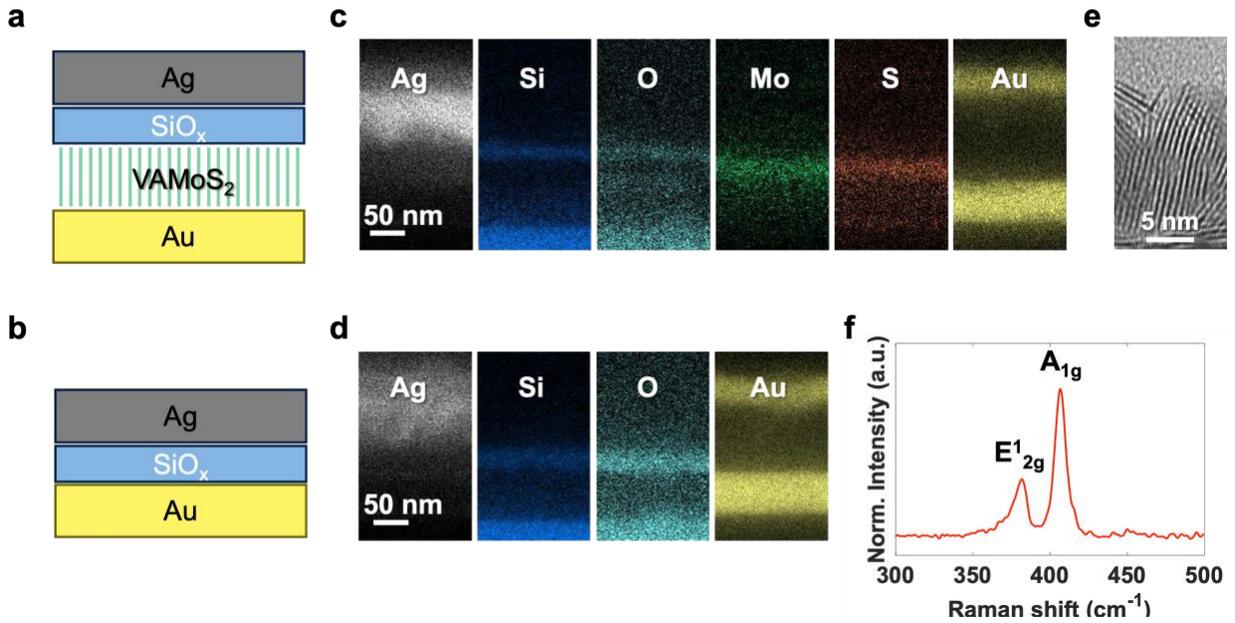

**Fig. 1.** Schematic cross-sections of the switching layers of (a) a $SiO_x/VAMoS_2$ and (b) a $SiO_x$-only device between the top (Ag) and bottom (Au) electrodes. EDXS elemental mapping confirming the presence of (c) Ag, Si, O, Mo, S, and Au in the $SiO_x/VAMoS_2$ device and (d) Ag, Si, O, and Au in the $SiO_x$-only device. (e) HRTEM cross-section revealing the vertically layered structure of $MoS_2$. (f) Raman spectrum of an as-grown $MoS_2$ film on an Au surface verifying the crystalline 2H-phase of $MoS_2$ with its characteristic $E^1_{2g}$ (382 cm$^{-1}$) and $A_{1g}$ (407 cm$^{-1}$) modes.

**Fig. 2.** Threshold Switching Characteristics of $SiO_x$/VAMoS$_2$ and $SiO_x$-only Devices.

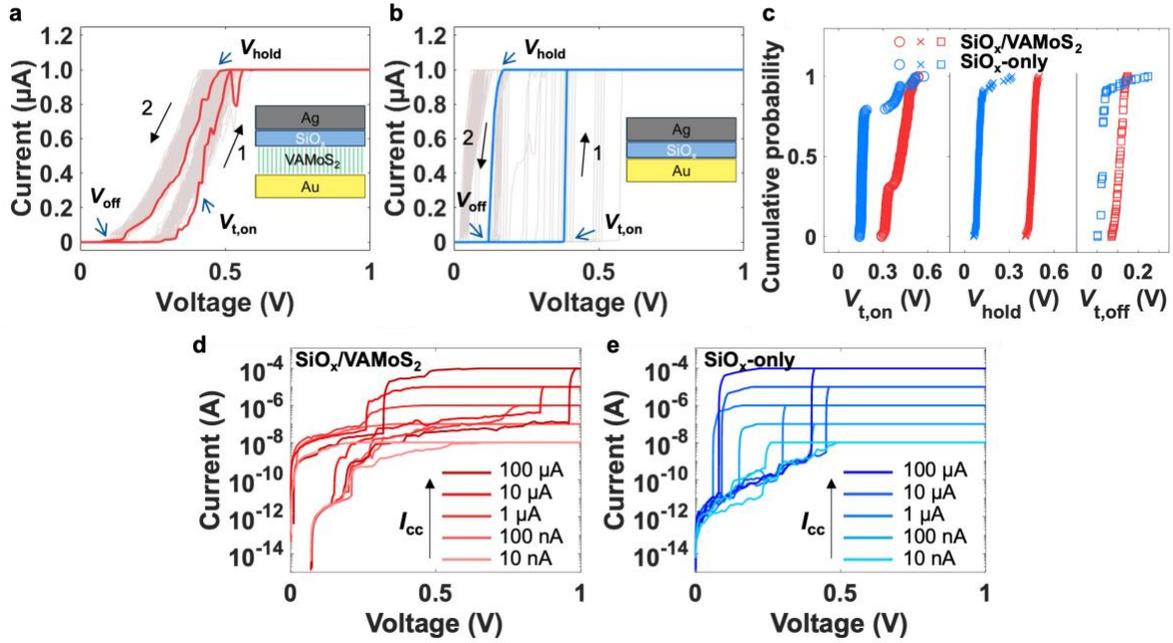

**Fig. 2.** (a) 128 consecutive dc $I$–$V$ sweeps for the $SiO_x$/VAMoS$_2$ device at $I_{cc}$ = 1 µA, showing a gradual current transition. The first sweep is marked in red. Inset: schematic cross-section of the $SiO_x$/VAMoS$_2$ device. (b) 128 consecutive $I$–$V$ sweeps for the $SiO_x$-only device, showing abrupt transitions. The first sweep is marked in blue. Inset: schematic cross-section of the $SiO_x$-only device. Characteristic voltages are indicated on the curves in (a) and (b). (c) Cumulative distribution functions of $V_{t,on}$, $V_{hold}$, and $V_{t,off}$ extracted from the data displayed in (a) and (b). Five subsequent $I$–$V$ sweeps under different $I_{cc}$ values (10 nA to 100 µA) showing the TS in (d) the $SiO_x$/VAMoS$_2$ device and (e) the $SiO_x$-only device.

**Fig. 3.** Switching dynamics of SiO$_x$/VAMoS$_2$ and SiO$_x$-only devices.

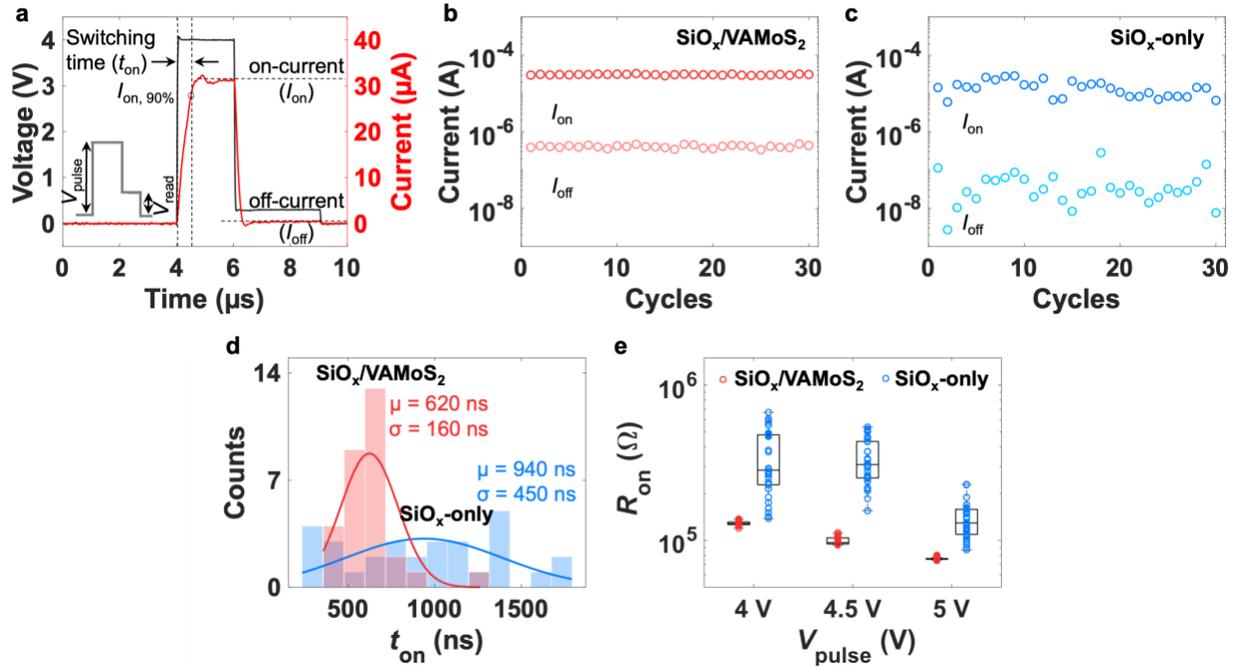

**Fig. 3.** (a) Representative transient response of a SiO$_x$/VAMoS$_2$ device under a voltage pulse of 4 V/2 μs, followed by a read pulse of 0.3 V/3 μs. The inset shows the applied waveform. On-current ($I_{on}$) and off-current ($I_{off}$) values extracted over 30 consecutive switching cycles for (b) a SiO$_x$/VAMoS$_2$ and (c) a SiO$_x$-only device. (d) Cycle-to-cycle variability of the switching time ($t_{on}$) over 30 pulses at 4 V, showing faster average and more reproducible switching in the SiO$_x$/VAMoS$_2$ device (μ = 620 ns, σ = 160 ns) than in the SiO$_x$-only device (μ = 940 ns, σ = 450 ns). (e) On-state resistance ($R_{on}$) as a function of pulse amplitude, with lower average values and reduced variability in the SiO$_x$/VAMoS$_2$ device than in the SiO$_x$-only device.

**Fig. 4.** Filament Formation and Variability Modeling in SiO$_x$/VAMoS$_2$ and SiO$_x$-only Devices.

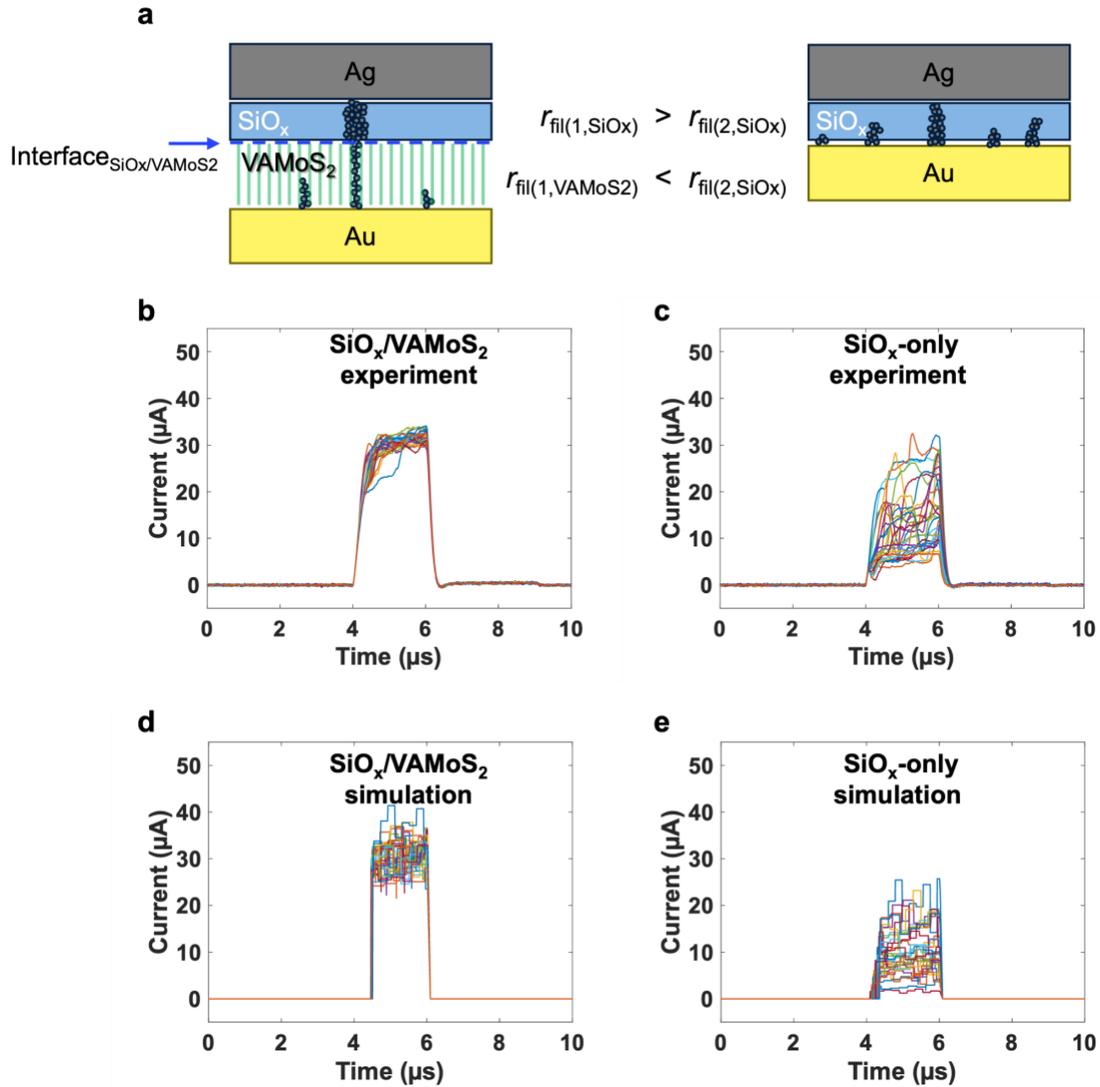

**Fig. 4.** (a) Schematic of conductive filament (CF) growth in the LRS: in SiO$_x$/VAMoS$_2$, vdW gaps guide ultrathin Ag seed CFs at the BE that extend into the SiO$_x$ layer, whereas in the SiO$_x$-only device, CFs nucleate randomly at the BE. Transient current responses under repeated 4 V/2 μs pulses over 30 cycles, showing faster and more consistent switching in (b) a SiO$_x$/VAMoS$_2$ device than in (c) a SiO$_x$-only device. Cycle-to-cycle variability fitting via the JART ECM model[42–44], reproducing the experimental data for (d) the SiO$_x$/VAMoS$_2$ and (e) the SiO$_x$-only devices.

## Supplementary Information

**Section S1. Cross-sectional TEM of Device Structures.**

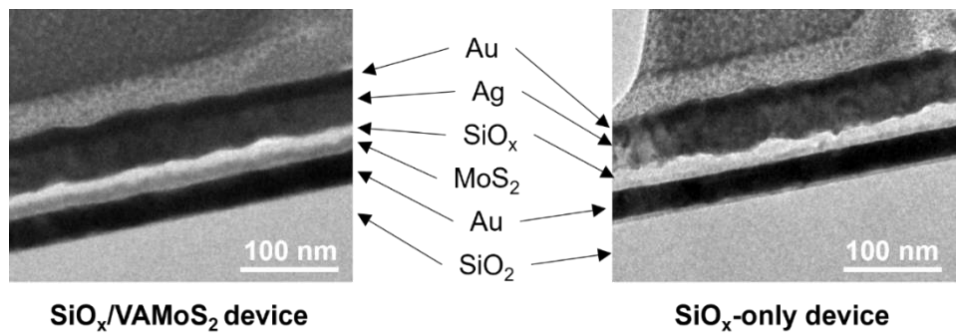

**Fig. S1.** Cross-sectional TEM images confirm the switching layers of the $SiO_x$/VAMoS$_2$ and $SiO_x$-only devices between the Ag TE and Au BE.

**Section S2. Distribution of Switching Voltage Parameters.**

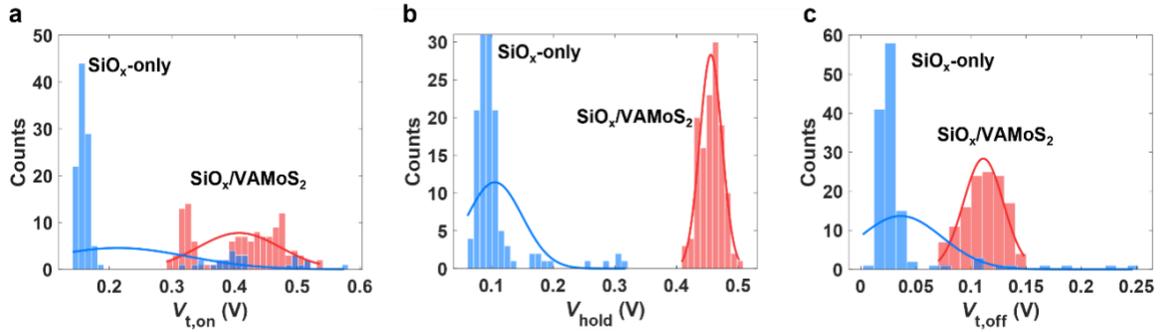

**Fig. S2.** Histograms of (a) on-threshold voltage ($V_{t,on}$), (b) hold voltage ($V_{hold}$), and (c) off-threshold voltage ($V_{t,off}$) extracted from 128 consecutive switching cycles for the $SiO_x/VAMoS_2$ and $SiO_x$-only devices.

**Section S3. Cycle-dependent Switching Voltages Parameters.**

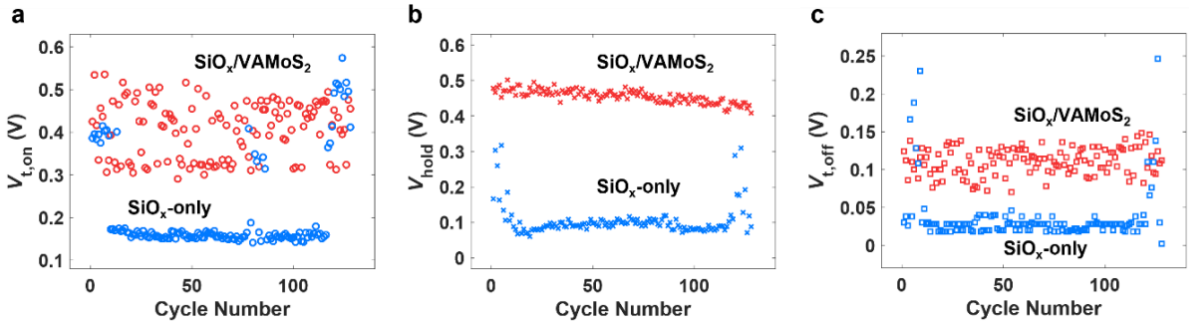

**Fig. S3.** Cycle-dependent switching voltages extracted from 128 consecutive cycles for the $SiO_x/VAMoS_2$ and $SiO_x$-only devices: (a) $V_{t,on}$, (b) $V_{hold}$, and (c) $V_{t,off}$.

**Section S4. Pulse Amplitude Dependence of Switching Time.**

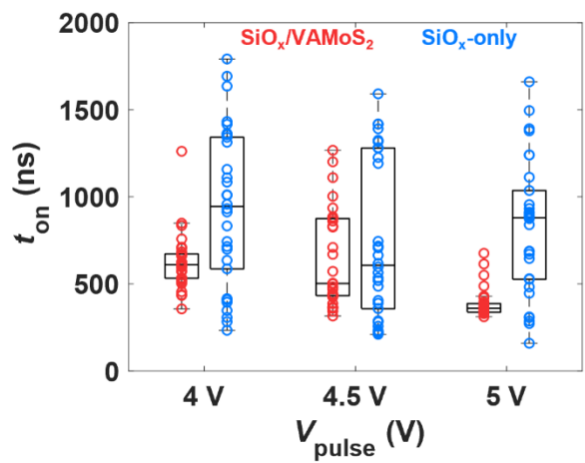

**Fig. S4.** Switching time ($t_{on}$) distributions measured from the consecutive cycles with voltage amplitudes of 4 V, 4.5 V, and 5 V, showing faster and more reproducible switching in the $SiO_x$/VAMoS$_2$ device.

## Section S5. Variability-aware Model.

**Table S1.** Model Parameters, Description and Values[1,2]

| Symbol/Parameter | Value | Symbol/Parameter | $SiO_x/VAMoS_2$ | $SiO_x$-only |
|---|---|---|---|---|
| $M_{me}$: molecular mass of silver | $1.7912 \cdot 10^{-25}$ kg | $r_{ac}$: radius of active electrode reaction area | 10 nm | 10 nm |
| $z$: charge transfer number | 1 | $r_{fil}$: filament radius | 1.22 nm | 0.8 nm |
| $\rho_{m,me}$: mass density of silver | $10.49 \cdot 10^3$ kg/m$^3$ | $r_{is}$: cross-sectional radius of ion hopping process | 2.4 nm | 2.0 nm |
| $m_r$: relative electron mass | 0.86 | $L$: length of switching/ oxide layer | 30 nm / 10 nm | 10 nm / 10 nm |
| $\Delta W_0$: eff. tunneling barrier height | 3.6 eV | $\rho_{fil}$: filament's electronic resistivity | $1.59 \cdot 10^{-8}$ Ω m | $1.59 \cdot 10^{-8}$ Ω m |
| $\alpha_{et}$: electron transfer coefficient | 0.1 | $R_{el}$: resistance of electrodes | 40 Ω | 40 Ω |
| $k_{0,et}$: electron transfer reaction rate | $3 \cdot 10^5$ m/s | $R_S$: series resistance | 40 Ω | 40 Ω |
| $c$: Silver ion concentration | $2 \cdot 10^{26}$ 1/m$^3$ | $T$: temperature | 298 K | 298 K |
| $\Delta G_{et}$: electron transfer activation barrier | 0.55 eV | $t_{0,nuc}$: prefactor of nucleation time | $2 \cdot 10^{-10}$ s | $2 \cdot 10^{-10}$ s |
| $a$: mean ion hopping distance | 0.2 nm | $\Delta G_{nuc}$: nucleation activation energy | 0.8 eV | 0.8 eV |
| $f$: ion hopping attempt frequency | $1 \cdot 10^{13}$ Hz | $N_c$: number of atoms to achieve stable nucleus | 3 | 3 |
| $\Delta G_{hop}$: ion migration barrier | 0.2 eV | $\alpha_{nuc}$: electron transfer coefficient for nucleation | 0.5 | 0.5 |
| $p_{jump}$: relative stochastic strength | 0 < Random number < 1 | $\Delta t_{step}$: simulation time step | 24.44 ps | 24.44 ps |
| $r_{Ag}$: radius of silver atom | 0.172 nm | $\delta$: strength of stochasticity | 0.0068 nm | 0.0159 nm |
| $d_{Ag}$: diameter of silver atom | 0.344 nm | $r_{fil,\sigma}$: standard deviation of Gaussian distribution | 0.24 nm | 0.24 nm |
| | | $r_{fil,trunc,low}$: lower boundary for truncated Gaussian distribution | 1.17 nm | 0.172 nm |
| | | $r_{fil,trunc,up}$: upper boundary for truncated Gaussian distribution | 1.27 nm | 0.85 nm |